\documentclass[usenatbib,babel]{iau}
\usepackage{graphicx}

\usepackage{amsmath}

\usepackage{xcolor}
\definecolor{Blue}{rgb}{0.25,0.41,0.88}

\title[Why do galactic spins flip in the cosmic web?] 
{Why do galactic spins flip in the cosmic web? \\ A Theory of Tidal Torques near saddles.}

\author[C. Pichon et al.]   
{Christophe Pichon$^{1}$\footnote{Email: {\tt pichon@iap.fr} },
 Sandrine Codis$^{1}$,
 Dmitry Pogosyan$^{2}$,\\
[\affilskip]
 Yohan Dubois$^{1}$, Vincent Desjacques$^{3}$
 \and Julien Devriendt$^{4}$
}
\affiliation{
$^{1}$ Institut d'Astrophysique de Paris \& UPMC, 98 bis Boulevard Arago, 75014, Paris, France \\%
$^{2}$ University of Alberta, 11322-89 Avenue, Edmonton, Alberta, T6G 2G7, Canada\\
$^{3}$  Universit\'e de Gen\`eve 24, quai Ernest Ansermet. 1211, Gen{\`e}ve, Switzerland\\
$^{4}$ Sub-department of Astrophysics, University of Oxford, Keble Road, Oxford OX1 3RH\\
}

\pubyear{2014}
\volume{308}  
\pagerange{119--126}
\jname{The Zeldovich Universe,
Genesis and Growth of the Cosmic Web}
\editors{Rien van de Weygaert, Sergei Shandarin, Enn Saar \& Jaan Einasto}
\begin{document}

\maketitle

\begin{abstract}
Filaments of the cosmic web drive spin acquisition of disc galaxies. The  
 point process of filament-type saddle represent best this environment and can be used to revisit the
Tidal Torque Theory  in the context of an anisotropic peak (saddle) background split.  
The constrained misalignment between the tidal tensor and the Hessian of the density field generated  in the vicinity of  filament saddle points
simply  explains  the corresponding transverse and 
longitudinal point-reflection symmetric geometry of spin distribution.
It predicts in particular an {\sl azimuthal} orientation of the spins of more massive galaxies
and spin {\sl alignment} with the filament for less massive galaxies.
Its scale dependence  also allows us to relate the transition mass corresponding to  the  alignment of dark matter halos' spin relative to the direction of  their neighboring filament to this 
geometry, and to predict accordingly it's  scaling with the mass of non linearity, as was measured in simulations.

\keywords{large-scale structure of universe, gravitational lensing, galaxies: statistics.
l}
\end{abstract}

\firstsection 

\section{Introduction}

\begin{figure}
\begin{center}
\includegraphics[scale=0.45]{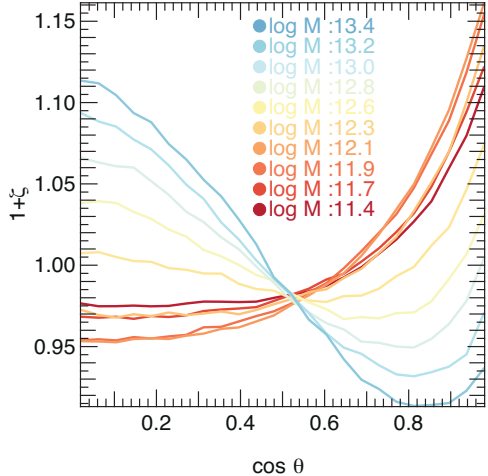}
\end{center}
\caption{The probability distribution of the cosine of the angle between the spin of dark haloes and the direction of the closest filament
as a function of mass in the Horizon simulation.  The probability to have  a small angle between the halo's spin and the filament's direction first{ \sl increases} as mass grows. At larger masses the  spin-filament alignment first  decays (at ${\cal M}^{\rm 2D}_{\rm tr}$), and then flips 
(at ${\cal M}^{\rm 3D}_{\rm tr}$) to predominately orthogonal orientations (from \cite{laigle2014}).
}
\label{fig:spin4pi}
\end{figure}

 Modern simulations based on  a well-established paradigm of cosmological structure formation predict a significant connection between the geometry and dynamics of the large-scale structure on the one hand, and the evolution of the physical properties of forming galaxies on the other. 
\cite{pichonetal11} have suggested that the large-scale coherence, inherited 
from the  low-density cosmic web, explains why cold flows are so efficient at producing thin high-redshift discs from the inside out.    They also predicted  that the distribution of the properties of galaxies measured relative to their cosmic web environment should reflect such a process. Both numerical (e.g. \cite{hahnetal07,codisetal12}, Fig~\ref{fig:spin4pi},  \cite{Libeskind13a}), and observational evidence (e.g. \cite{tempel13}) have recently supported this scenario. 
In parallel, much observational effort has been invested to control the level of intrinsic alignments of galaxies as a potential source of systematic errors in weak gravitational lensing measurements (e.g. \cite{HRH00}). 
It is therefore of interest to explain from first principles why such intrinsic alignments arise,
so as to possibly temper their effects (see also \cite{codis14}).  

Yet, understanding the effect of this cosmic anisotropy on galactic morphology is a challenging task. 
The difficulty is two-fold: 
i) the geometry of the flow within filaments  is complex: the spin distribution
 is intrinsically point-reflection symmetric relative to saddles,  and confined to filaments
 ii) the cosmic web itself is strongly anisotropic and multiscale.
In this paper, we will try and address these challenges and formalize the corresponding theory of anisotropic secondary infall.
 Specifically, we will model the intrinsically 3D geometry of galactic accretion while taking into account the geometry of the tidal and density field near a {\sl typical}  ÒsaddleÓ point.
Indeed, saddle points  define an  {\sl point process} which accounts for the presence of 
 filaments embedded in walls, two critical ingredient in shaping the spins of galaxies.
  A proper account of the anisotropy of the  environment in this context will allow us to demonstrate why, as measured in simulations, the spin of the forming ({\sl low mass}) galaxies  are first aligned with the filamentÕs direction with a quadratric point symmetric geometry (Fig~\ref{fig:spin4pi} and \cite{laigle2014}).
 While relying on a straightforward extension of Press Schechter's theory, 
 we will also demonstrate that  {\sl massive} galaxies will have their spin preferentially along 
 the azimuthal direction, and predict the corresponding scaling of the spin-flip transition mass
 with the (redshift dependent) mass of non-linearity, on the basis of the so-called cloud-in-cloud problem,
 applied at the peak (filamentary) background split level.
 %
 \begin{figure}
\begin{center}
\includegraphics[width=0.35\columnwidth]{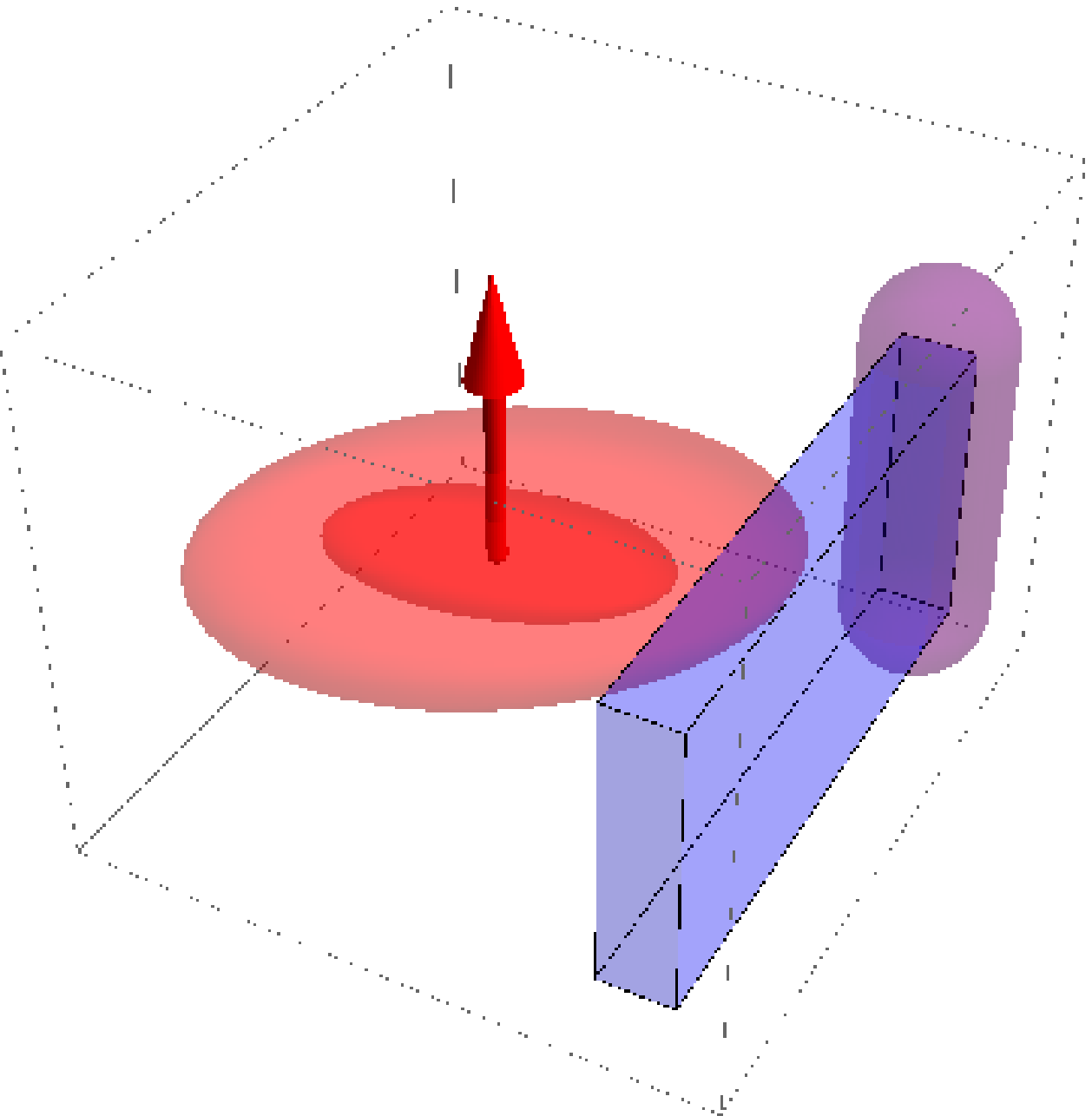} \hskip 1cm
\includegraphics[width=0.35\columnwidth]{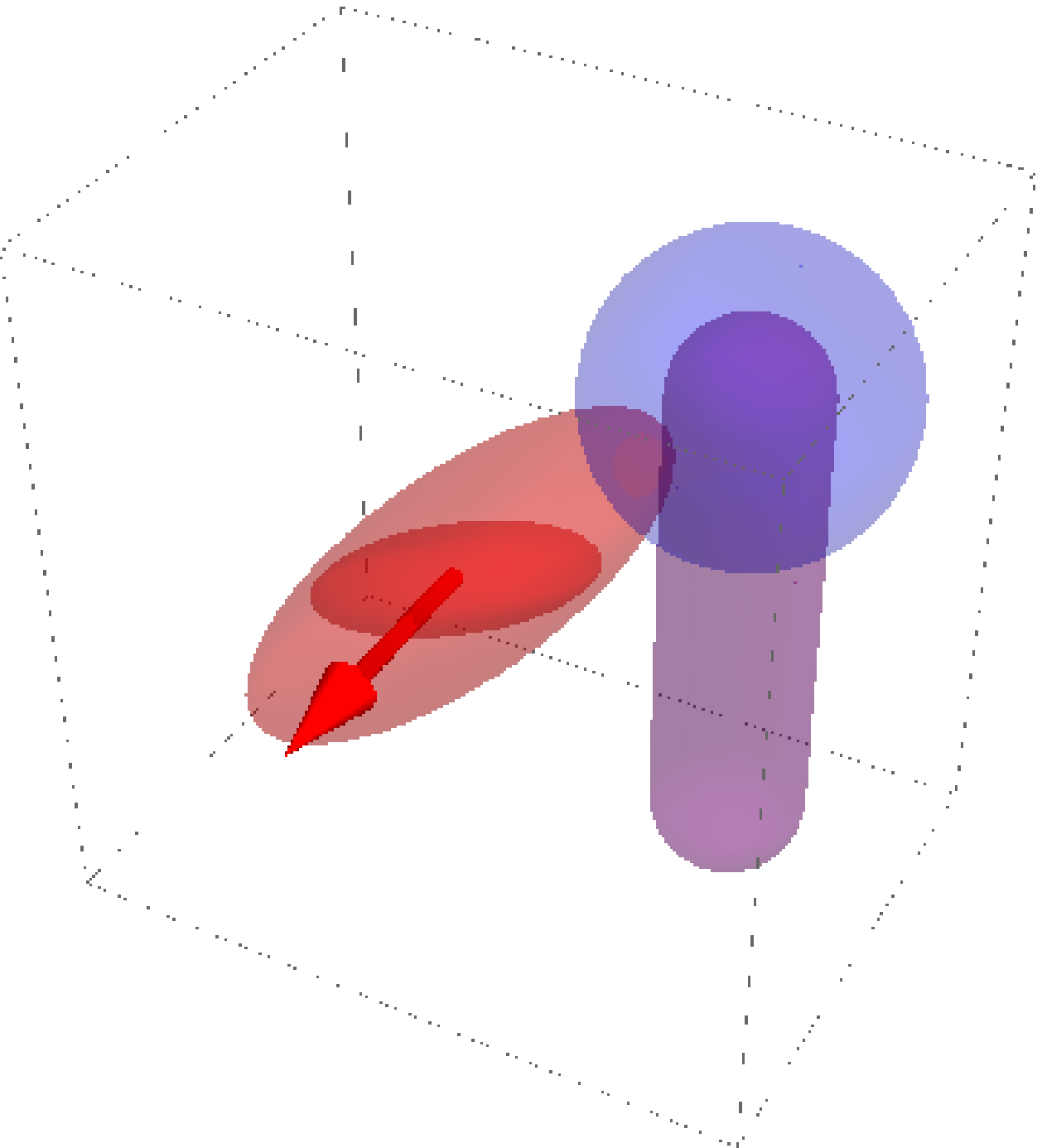}
\caption{sketch of main differential alignment between hessian and tidal responsible for $\mathbf e_z$ and $\mathbf e_\phi$
component of spin.
{\sl Left:} the two tensors in light and dark red, are misaligned 
as they feel differently the neighboring wall (blue) and filament (purple), inducing a spin
parallel to the filament (red arrow). {\sl Right:} correspondingly, the differential pull from the filament 
(purple)
and the density gradient towards the peak (blue) generates a spin (red arrow) along the azimuthal direction. 
\label{principe}}
\end{center}
\end{figure}
 
Qualitatively, the idea is the following: given a triaxial  saddle constraint,  the  misalignment between the tidal tensor and the Hessian of the density field   simply  explains  the transverse and  longitudinal point-reflection symmetric geometry of spin distribution  in their vicinity.
It arises because the two tensors probe different scales: given their relative correlation lengths,
 the Hessian probes more directly its closest neighborhood, while the tidal field, 
somewhat larger scales, see Fig.~\ref{principe}.
Within the plane perpendicular to the filament axis at the saddle point, the dominant wall (corresponding to the longer axis of the cross section of the saddle point) will  re-orient more the Hessian than the tidal field, which also feels the denser, but typically further away saddle point.
This net misalignment will induces spin perpendicular to that plane i.e along the filament. 
This effect will produce a quadru-polar, point reflection symmetric distribution of the polar component of the spin which will  be strongest at some four points,  not far off axis. 
Beyond a couple of correlation lengths away from those  four points, the effect of the tidal field induced by the saddle point will subside, as both tensor become
more spherical.  
Conversely, in  planes perpendicular to that plane,  e.g.  containing the dominant wall and the filament,  a similar process will misalign both tensors. This time, the two anisotropic features differentially pulling the tensors are the filament on the one hand, and the density gradient towards  the peak on the other.  The net effect of the corresponding misalignment will be to also spin up halos perpendicular to that plane,  along the azimuthal direction. By symmetry, an  anti-clockwise tidal spin will be generated on the other side of the saddle point.
\par
\begin{figure}
\begin{center}
\includegraphics[width=0.55\columnwidth]{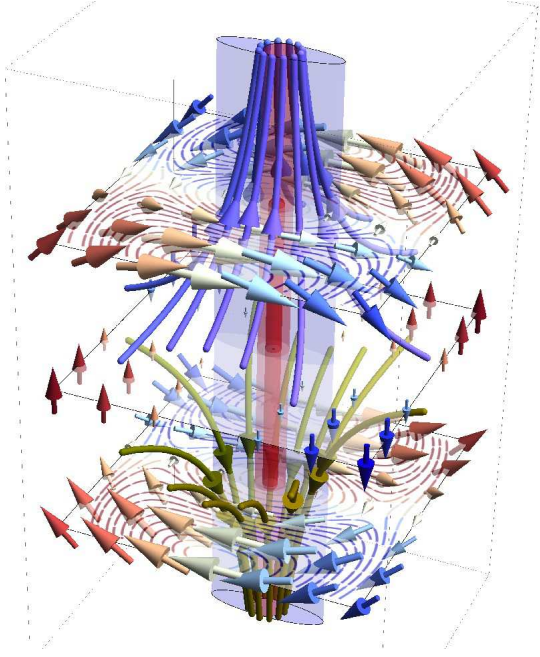}
\caption{\small{
The velocity and Spin flow near a vertical filament (in red) embedded in 
a (purple) wall. The purple and green flow lines trace the (Lagrangian) 3D velocities
(upwards and downwards respectively).
The red and blue arrows show
the spin 3D distribution, while 
the three horizontal cross sections show spin flow lines in the corresponding plane.
Note that the  spin is along $\mathbf e_z$ in the mid plane and along $\mathbf e_\phi$ away from it,
and that it rotates in opposite direction above and below the mid-plane.
See also {\tt http://www.iap.fr/users/pichon/AM-near-saddle.html}
}}
\label{fig-momentum3D}
\end{center}
\end{figure}

Hence, as the theory developed below will allow us to predict, the geometry of spin near filament-saddle points is the following: it is aligned with the filament in the median plane (within four anti-symmetric quadrants), and aligned with the azimuthal direction away from that plane, see Fig.~\ref{fig-momentum3D}.
The stronger the  triaxiality the stronger the amplitude.  Conversely, if the saddle point becomes degenerate in one or two directions, the component of the spin in the corresponding direction will vanish. For instance, a saddle point in the middle of  a very long filament will only display alignment with that filament axis, with no azimuthal component.

The paper is organised as follows. 
Section~\ref{sec:2D} presents the expected  Lagrangian  spin distribution near filaments, assuming cylindrical symmetry, while 
Section~\ref{sec:3D}  revisits this distribution in three dimensions
for realistic typical 3D saddle points.

\section{ Spin in cylindrical symmetry   }
\label{sec:2D}

Let us  start while assuming that the filament is of infinite extent, so that we can restrict ourselves to cylindrical symmetry in two dimensions.
This is of interest as the spin is then along the filament axis by symmetry and its derivation in the context of Tidal Torque theory (TTT) is much simpler.
It captures already in part the mass transition, as 
we can define the mean extension of a given quadrant of spin with a given 
polarity.


Under the assumption that the {\sl  direction} of  the spin 
 along the $z$ direction is well represented by the 
  anti-symmetric (Levi Civita) contraction 
of the tidal field and hessian (e.g. \cite{Schafer2012}),
it becomes a quadratic function of the second and fourth derivatives of the potential.
As such, it becomes possible to compute expectations of it subject to its relative position
to a peak with a given geometry (which would correspond to the cross section of the filament in that plane).
In contrast, standard TTT  relies, more correctly, on the  inertia tensor in place of the Hessian.
Even though they have inverse curvature of each other, their set of eigen-directions are locally the same, 
 so we expect the induced spin {\sl direction}-- which is the focus of this paper, to be the same,
 so long as the inertia tensor is well described by its local Taylor expansion.

Any matrix of second derivatives $f_{ij}$ -- rescaled so that $\left\langle (\Delta f)^{2} \right\rangle=1$-- can be decomposed into its trace $\Delta f$, and its detraced components in the frame of the separation $f^{+}=(f_{11}-f_{22})/2$ and $f^{\times}=f_{12}$. Then all the correlations between two such matrices, $f_{ij}$ and $g_{ij}$ can be decomposed irreducibly as follows.
Let us call $\xi_{f g}^{\Delta\Delta}$ ,
$\xi_{f g}^{\Delta+}$ and $\xi_{f g}^{\times\times}$ 
the correlation functions in the frame of the separation (which is the first coordinate here) between the second derivatives of the field $f$ and $g$ separated by a distance $r$:
\begin{equation}
\xi_{f g}^{\Delta\Delta}(r)=\left\langle\Delta f \Delta g\right\rangle,\,\,\,\,
\xi_{f g}^{\Delta +}(r)=\left\langle\Delta f g^{+} \right\rangle,\,\,\,\,
\xi_{f g}^{\times\times}(r)=\left\langle f^{\times}\!g^{\times}\right\rangle\,. \nonumber 
\end{equation}
All other correlations are trivially expressed in terms of the above as 
$
\left\langle f^{\times} \Delta g\right\rangle=0,$ $
\left\langle f^{+}\! g^{\times} \right\rangle=0,$ $
\left\langle f^{+}\! g^{+} \right\rangle=\frac 1 4 \xi_{f g}^{\Delta\Delta}(r)-\xi_{f g}^{\times\times}(r)\nonumber\,.
$
Here, we consider two such fields, namely the gravitational potential $\phi$ and the density contrast $\delta$. In the following these two fields and their first and second derivatives are assumed to be rescaled by their variance $\sigma_{0}^{2}=\left\langle \phi^{2}\right\rangle$, $\sigma_{1}^{2}=\left\langle  (\nabla\phi)^{2}\right\rangle$, $\sigma_{2}^{2}=\left\langle  (\delta=\Delta\phi)^{2}\right\rangle$, $\sigma_{3}^{2}=\left\langle  (\nabla\delta)^{2}\right\rangle$ and $\sigma_{4}^{2}=\left\langle  (\Delta\delta)^{2}\right\rangle$. The shape parameter is defined as $\gamma=\sigma_{3}^{2}/(\sigma_{2}\sigma_{4})$.

The Gaussian joint PDF of the gravitational field, its first and second derivatives and the first and second derivatives of the density is sufficient to compute the expectation of any quantity involving  derivatives of
 the potential  and the density up to second order.
The two-point covariance matrix can be  derived from the power spectrum of the potential, the result being a function of the above defined nine functions (for $fg=\phi\phi,\phi \delta,\delta\delta$).
Once the joint PDF is known, it is straightforward to compute conditional PDFs. Simple 
algebra yield the conditional density and spin as a function of separation and geometry of 
the saddle.
In details, given a contrast $\nu$ and  a geometry for the saddle defined by $\kappa=\lambda_{1}-\lambda_{2},I_{1}=\lambda_{1}+\lambda_{2}$ (where $\lambda_{1}>\lambda_{2}$ are the two eigenvalues of the Hessian of the density field $\mathbf{ H}$ -- both negative for a peak),
 the mean density contrast, $\langle \delta  | {\rm ext} \rangle$
 (in units of $\sigma_{2}$) around the corresponding extremum can be computed as 
\begin{align}
\delta(\mathbf{r},\kappa,I_{1},\nu| {\rm ext})=&\frac{I_{1} (\xi_{\phi \delta}^{\Delta\Delta} +\gamma \xi_{\phi\phi}^{\Delta\Delta} )+
\nu  (\xi_{\phi \phi}^{\Delta\Delta}+\gamma \xi_{\phi \delta}^{\Delta\Delta})}{1-\gamma ^2} 
+4\left(\mathbf{\hat r}^{\rm T} \!\!\cdot \overline{\mathbf{ H}}\cdot\mathbf{\hat r}\right) \xi_{\phi \delta}^{\Delta +} \,,
\end{align}
where $\overline{\mathbf{ H}}$ is the detraced Hessian of the density and
$\mathbf{\hat r}={\mathbf{r} }/{r}$
so that 
$
\mathbf{\hat r}^{\rm T}\!\! \cdot \overline{\mathbf{ H}}\cdot\mathbf{\hat r}= \kappa {\cos (2 \theta )}/{2}\,,
$ with
$r$ is the distance to the extremum and $\theta$ the angle from the eigen-direction corresponding to the first eigenvalue $\lambda_{1}$ of the extremum. When $r$ goes to zero,  given the properties of the $\xi$ functions,
the density trivially converges to the constraint $\nu$.

In 2D, the (rescaled) spin is a  scalar  given by
$
{L}_z (\mathbf r) = \varepsilon_{ij}  \phi_{il}   x_{jl}\,, \label{eq:defL2D}
$
where   $\epsilon$ is the rank 2 Levi-Civita tensor.
Hence the spin generated by TTT as a function of the polar
position, $(r,\theta)$  subject to the same extrema constraint at the origin with 
contrast,
$\nu$, and principal curvatures $(\lambda_{1},\lambda_{2})$
is given by the sum of a quadrupole ($\propto \sin 2\theta$) and an octupole ($\propto \sin 4\theta$,
since   
 $\mathbf{\hat r}^{\rm T}\!\!\cdot \epsilon \cdot \overline{\mathbf{ H}}\cdot\mathbf{\hat r}= 
-\kappa{\sin (2 \theta )}/{2} $):
\begin{equation}
\langle {L}_z  | {\rm ext}\rangle=   L_z(\mathbf{r} ,\kappa,I_{1},\nu| {\rm ext}) 
 =-16(\mathbf{\hat r}^{\rm T}\!\!\cdot \epsilon \cdot \overline{\mathbf{ H}}\cdot\mathbf{\hat r})\,
 \Big(L_{z}^{(1)}(r)+2 (\mathbf{\hat r}^{\rm T}\!\! \cdot \overline{\mathbf{ H}}\cdot\mathbf{\hat r}
) L_{z}^{(2)}(r)\Big)    \label{eq:defL2D}
\,,
\end{equation}
 where  the octupolar component $L_{z}^{(2)}$ can be written as
$
  L_{z}^{(2)}(r)=  (\xi_{\phi x}^{\Delta\Delta}\xi_{\delta \delta}^{\times\times}-\xi_{\phi \delta}^{\times\times}\xi_{\delta \delta}^{\Delta\Delta}
  )
$,
     and the quadrupolar coefficient
     $L_{z}^{(1)}(r)$ reads   
     \begin{align}	
  L_{z}^{(1)}(r)=& 
  \frac{\nu}{1-\gamma^{2}}
   \left[
  (\xi_{\phi \phi}^{\Delta +}+\gamma \xi_{\phi \delta}^{\Delta +})\xi_{\delta\delta}^{\times\times}
  -(\xi_{\phi \delta}^{\Delta +}+\gamma \xi_{\delta\delta}^{\Delta +})\xi_{\phi \delta}^{\times\times}
  \right]\nonumber
  \\
 &+\frac{ I_{1}}{1-\gamma^{2}} \left[
 (\xi_{\phi \delta}^{\Delta +}+\gamma \xi_{\phi \phi}^{\Delta +})\xi_{\delta\delta}^{\times\times}
-
  (\xi_{\delta\delta}^{\Delta +}+\gamma \xi_{\phi \delta}^{\Delta +})\xi_{\phi \delta}^{\times\times}
    \right]\,.
\nonumber     \end{align}
Eq.~(\ref{eq:defL2D}) is remarkably simple.
  As expected, the spin, $L_z$, is identically null if the filament is axially symmetric  ($\kappa=0$).
It is zero along the principal axis of the Hessian (where $\theta=0 \mod \pi/2$ for which $\mathbf{\hat r}^{\rm T}\cdot \epsilon \cdot \mathbf{\hat r}=0$).
Near the peak, the anti-symmetric, $\sin(2\theta)$,  component dominates, and the spin distribution is quadripolar (see the midplane of
Fig~\ref{fig-momentum3D}). 

\begin{figure}
\begin{center}
{\center\includegraphics[width=0.65\columnwidth]{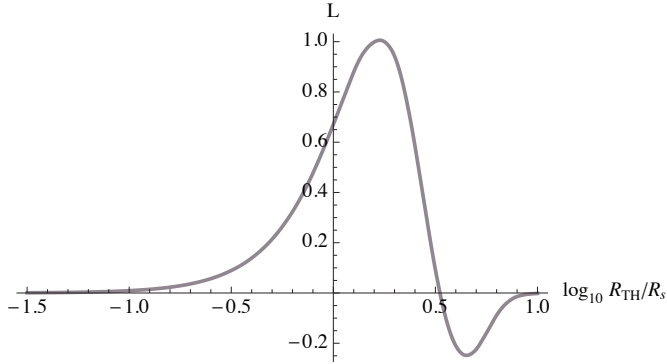}}
\caption{Evolution of the amount of algebraic 2D spin in sphere of radius $R_{\rm TH}$ centered on $r_{\star}$. The density power spectrum index is $n=-3/2$, the height of the peak in $(0,0)$ is $\nu=1$ and principal curvatures $\lambda_{1}=-1,\lambda_{2}=-2$. The amplitude of the spin is normalised by its maximum value around $R_{\rm TH}=r_{\star}$.}
\label{fig:spinTH}
\end{center}
\end{figure}

Let us now understand how much spin is contained within spheres of increasing radius that would feed the forming object at different stage of its evolution.
 For instance let us assume there is a small-scale overdensity at (one of the four) location of maximum spin (denoted $r_{\star}$ hereafter) and let us filter the spin field with a top-hat window function centered on $r_{\star}$ and of radius $R_{\rm TH}$. The resulting amount of spin as a function of this top-hat scale is displayed in Fig.~\ref{fig:spinTH}. During the first stage of evolution, the central object will acquire spin constructively until it reaches a Lagrangian size of radius $R_{\rm TH}=r_{\star}$ and feels the two neighbouring quadrants of opposite spin direction. The spin amplitude then decreases and becomes even negative before it is fed by the last quadrant of positive spin. The minimum is reached for radius around $2.4 r_{\star}$. This result does not change much with the contrast and the geometry of the peak constraint.
 %
%
Let us now predict the mass that corresponds to maximum spin i.e. the mass contained in a sphere of radius $R_{\star}$. First, let us compute $r_\star$,   as the radius for which $L_z(r,\theta=\pi/4)$
is maximal as a function of $r$. Indeed, for small enough $\kappa$,
   the quadruple dominates, and the extremum is along $\theta=\pi/4$.
The area of a typical quadrant, in which the spin has the same polarity, can then simply be expressed as  
$
{\cal A}=\ \pi {\lambda_{2}}/{\lambda_{1}} (2 r_{\star})^{2  }/4\,, \label{eq:defarea}
$
where $\lambda_{1}<\lambda_{2}< 0$ are the two eigenvalues of the Hessian and 
 $r_{\star}=r_{\star}(\nu,\kappa)$ is the position of a maximum of spin from the peak.
 {
Because of the quadrupolar anti-symmetric geometry of the spin distribution near the saddle point,
it is typically twice as small (in units of the smoothing length) as one would naively expect.}

With prior knowledge of the distribution of the shape, $\kappa$ and height, $\nu$ for 2D peaks  and of the maximal area, $\cal{A}(\kappa,\nu)$, corresponding to spins with the same polarity, we may {\sl define} the transition  corresponding mass as  (with $\Sigma_{0}$  the cosmic mean surface density)
\begin{equation}
{\cal M}^{\rm 2D}_{\rm tr }(L_{s})=\Sigma_0 \int d\nu\,d \kappa
 {\cal A}(\nu,\kappa){\cal P}(\nu,\kappa|{\rm pk})\,,
\label{eq:Mtr2D}
\end{equation}
 Following \cite{pogo09}, it is straightforward to derive this PDF, ${\cal P}$, for a {\sl peak} to have height $\nu$ and geometry $\kappa,I_{1}$  so that
\begin{multline}
 {\cal P}(\nu,\kappa,I_{1}|{\rm pk})=\frac{\sqrt{3}\kappa|(I_{1}-\kappa)(I_{1}+\kappa)|}{2\pi\sqrt{1-\gamma^{2}}}\Theta(-\kappa-I_{1})
 \exp \left(\!\!-\frac 1 2 \!\left(\frac{\nu+\gamma I_{1}}{\sqrt{1-\gamma^{2}}}\right)^{\!2}\!\!\!\!-\frac 1 2 I_{1}^{2}\!\!-\kappa^{2}\!\!\right)
\,. \nonumber
 \end{multline}

Given the shape of  $\cal P$ near it maximum, we can approximate ${\cal M}^{\rm 2D}_{\rm tr }$
in Eq.~(\ref{eq:Mtr2D}) as 
\begin{equation}
{\cal M}^{\rm 2D}_{\rm tr }(z)= {\Delta}N
 \frac{\lambda_{2,\star}}{\lambda_{1,\star}} \left( \frac{r_{\star}}{L_s}\right)^{2}\,
  M_s(z) \equiv \alpha {M}_{s}\,,
\label{eq:Mtr2Dnew}
\end{equation}
where ${\Delta }N={\cal P} (\nu_\star,\kappa_{\star}|{\rm pk}) \Delta \nu \Delta \kappa$,
and $M_s(z) \equiv \pi L_s^2(z) \Sigma_0 $. Here the $\lambda$'s and $\nu$ are evaluated at the maximum of ${\cal P} $
and the $ \Delta$'s represent the local inverse curvature at the peak of that distribution.
For scale invariant power spectra, the calculation shows that  $\alpha(n=-1)\sim 1/11$.
This is one of the main results of this investigation.
It states that, in the framework of anisotropic peak background split of TTT 
near a typical saddle point (for a GRF of density index $\sim-1$), the transition mass is {\sl predicted} to be smaller  than the scaling mass, $M_s$, by an order of magnitude. This is what \cite{codisetal12} found while analyzing the scaling of the transition mass with the mass of non-linearity (see Fig~\ref{fig:spin4pi}).
%
\section{ The 3D spin near and along filaments }
\label{sec:3D}

Let us now turn to the truly three dimensional theory of tidal torques 
in the vicinity of a typical filament saddle point,  see Fig.~\ref{fig-momentum3D}.
The main motivation is that 
the 3D saddle geometry  captures fully the second (spin flip) mass transition.
In three dimensions, we must consider two competing processes.
If we vary the  radius corresponding to the Lagrangian patch centered on the 
running point,
 we have a spin-up (along $\mathbf e_z$) arising from  the running point to wall-running point to saddle tidal misalignment and 
 a second spin-up  (along $\mathbf e_\phi$) arising from running point to filament- running point to peak tidal misalignment.

 In order to compute the spin distribution, the formalism developed in Section~\ref{sec:2D} is easily  extended to 3D. A critical (including saddle condition) point constraint is imposed. 
 The resulting mean density field subject to that constraint becomes
 (in units of $\sigma_{2}$):
\begin{multline}
\delta(\mathbf{r},\kappa,I_{1},\nu| {\rm ext})=
\frac{I_{1} (\xi_{\phi\delta}^{\Delta\Delta} +\gamma \xi_{\phi\phi}^{\Delta\Delta} )}{1-\gamma ^2}+
\frac{\nu  (\xi_{\phi \phi}^{\Delta\Delta}+\gamma \xi_{\phi \delta}^{\Delta\Delta})}{1-\gamma ^2}
+\frac {15}{2}\left(\mathbf{\hat r}^{\rm T} \cdot \overline{\mathbf{ H}}\cdot\mathbf{\hat r}\right) \xi_{\phi \delta}^{\Delta +} \,,
\end{multline}
where again $\overline{\mathbf{ H}}$ is the {\sl detraced} Hessian of the density and
$\mathbf{\hat r}={\mathbf{r} }/{r}$
and we define in 3D $\xi_{\phi x}^{\Delta +} $ as
$
\xi_{\phi \delta}^{\Delta +} =\left\langle\Delta \delta,\phi^{+}\right \rangle,
$
with 
$\phi^{+}=\phi_{11}-(\phi_{22}+\phi_{33})/2$. Note that $  \hat {\mathbf r}^{\rm T}\cdot \overline{\mathbf{ H}}\cdot \hat {\mathbf{r} }$
is a  scalar quantity defined explicitly as
     $  \hat { r}^{}_{i}\overline{{ H}}_{ij}\hat {{r} }_{j}\,.$
 As in 2D, the  expected spin  can also be computed. 
 In 3D, the spin is a  vector, which    components are  given by
$
{L}_i = \varepsilon_{ijk}  \delta_{kl}\phi_{lj} \,, \label{eq:defL3D}
$
with 
$\boldsymbol\epsilon$ the rank 3 Levi Civita tensor.
 It is found to be orthogonal to the separation and can be written as the sum of two terms
 \begin{equation}
 \mathbf{L}(\mathbf{r},\kappa,I_{1},\nu| {\rm ext})=-15
\Big(\mathbf L^{(1)}(r)
+\mathbf L^{(2)}(\mathbf{r})\Big)\cdot(\hat {\mathbf r}^{\rm T}\!\!\cdot \boldsymbol\epsilon\cdot \overline {\mathbf{ H}}\cdot \hat {\mathbf{r} })
\,, \label{eq:defL3Dsol}
\end{equation}
where 
  $\mathbf L^{(1)}$ depends on  height, $\nu$, and on the trace of the Hessian $I_{1}$ but not on  orientation 
     \begin{align}	
 \mathbf{ L}^{(1)}(r)=&\left(  \frac{\nu}{1-\gamma^{2}}  \left[
  (\xi_{\phi \phi}^{\Delta +}+\gamma \xi_{\phi \delta}^{\Delta +})\xi_{\delta\delta}^{\times\times}
  -(\xi_{\phi \delta}^{\Delta +}+\gamma \xi_{\delta\delta}^{\Delta +})\xi_{\phi \delta}^{\times\times}
  \right]\right.\nonumber
  \\
 & + \left.
 \frac{I_{1}}{1-\gamma^{2}}
  \left[
 (\xi_{\phi \delta}^{\Delta +}+\gamma \xi_{\phi \phi}^{\Delta +})\xi_{\delta\delta}^{\times\times}
-
  (\xi_{\delta\delta}^{\Delta +}+\gamma \xi_{\phi \delta}^{\Delta +})\xi_{\phi \delta}^{\times\times}
    \right]\,\right) \mathbb{I}_{3}\,,\nonumber
     \end{align}
     and $\mathrm L^{(2)}(\mathbf{r})$ now depends on $\overline{\mathbf{ H}}$ and on  orientation:
          \begin{multline}	
\mathbf{L}^{(2)}(\mathbf{r})=
  -\frac{5}{8}
   \left[
 2((\xi_{\phi \delta}^{\Delta +}-\xi_{\phi \delta}^{\Delta \Delta})\xi_{\delta\delta}^{\times\times}
  -( \xi_{\delta\delta}^{\Delta +}-\xi_{\delta\delta}^{\Delta \Delta})\xi_{\phi \delta}^{\times\times}) 
  \overline{\mathbf{ H}}
 \right.\nonumber
  \\
+  \left.
( (7\xi_{\delta\delta}^{\Delta \Delta}+5 \xi_{\delta\delta}^{\Delta +})\xi_{\phi \delta}^{\times\times}-(7\xi_{\phi \delta}^{\Delta \Delta}+5\xi_{\phi \delta}^{\Delta +})\xi_{\delta\delta}^{\times\times}
  )
(\hat {\mathbf r}^{\rm T}\cdot  \overline{\mathbf{ H}}\cdot \hat {\mathbf{r} })\mathbb{I}_{3}
    \right]\,,
   \nonumber
     \end{multline}
(with    $\mathbb{I}_{3}$ the identity matrix)  operating on the  {\sl vector}
$
    \left( \hat {\mathbf r}^{\rm T}\!\!\cdot \boldsymbol\epsilon\cdot \overline{\mathbf{ H}}\cdot \hat {\mathbf{r} }\right)_j=    \hat {\mathbf r}^{}_{i}\epsilon_{ijk}\overline{\mathbf{ H}}_{kl}\hat {\mathbf{r} }_{l}\,.
$
     Note that all the dependence with the distance $r$ is encoded in the $\xi$ functions,
      while the geometry of the critical point  is encoded in the terms corresponding to the peak height, trace and detraced part of the Hessian, while the orientation of the separation is encoded in $\mathbf{\hat r}$.
     Eq.~(\ref{eq:defL3Dsol}) is also remarkably simple: 
     as expected the symmetry of the model induces zero spin along the principal directions of the Hessian (where $\hat {\mathbf r}^{\rm T}\!\!\cdot \epsilon\cdot \overline{\mathbf{ H}}\cdot \hat {\mathbf{r} }=0$) and a point reflection symmetry ($\mathbf{\hat r}\rightarrow-\mathbf{\hat r} $),
     see Fig.~\ref{fig-momentum3D}.


Let us now compute the mean values of $\nu$, $\lambda_{1}<\lambda_{2}<0<\lambda_{3}$ of a typical   filament-type saddle-point.
Starting from the so-called Doroskevich formula for the PDF:
\begin{multline}
{\cal P}(\nu,\lambda_{i})=
\!
\frac{135 \left( 5 /{2\pi} \right)^{3/2}
}{4\sqrt{1-\gamma^{2}}}
\!
\exp 
\!\left[-\frac 1 2 \zeta^{2}-3 I_{1}^{2}+\frac {15} 2 I_{2}\right]
\times (\lambda_{3}-\lambda_{1})(\lambda_{3}-\lambda_{2})(\lambda_{2}-\lambda_{1})\,,\nonumber
\end{multline} 
where
$\zeta=(\nu+\gamma I_{1})/\sqrt{1-\gamma^{2}}$,
$I_{1}=\lambda_{1}+\lambda_{2}+\lambda_{3}$,
$I_{2}=\lambda_{1}\lambda_{2}+\lambda_{2}\lambda_{3}+\lambda_{1}\lambda_{3}$
and $I_{3}=\lambda_{1}\lambda_{2}\lambda_{3}$,
subject to the constraint, this PDF becomes
\begin{equation}
{\cal P}(\nu,\lambda_{i}|\textrm{skl})=
 \frac{26460 \sqrt{5 \pi } {\cal P}(\nu,\lambda_{i}) I_{3} \Theta(\lambda_{3})
}{1421 \sqrt{2}-735 \sqrt{3}+66 \sqrt{42}}\Theta(-\lambda_{2}-\lambda_{3})\,,\label{eq:defskelsaddle}
\end{equation} 
after imposing the condition of saddle point $|\det x_{ij}|\delta_{D}({x_i})\Theta(\lambda_{3})\Theta(-\lambda_{2})$ and the
 additional constraint of a skeleton-like saddle, which is $\lambda_{2}+\lambda_{3}<0$. 
The expected value of the density and the eigenvalues at a skeleton saddle position reads $\left\langle\nu\right\rangle\approx 1.25 \gamma$, $\left\langle\lambda_{1}\right\rangle\approx-1.0$, $\left\langle\lambda_{2}\right\rangle\approx -0.56$ and $\left\langle\lambda_{3}\right\rangle\approx 0.31$.


The transition mass, ${\cal M}^{\rm 3D}_{\rm tr}$ may then be defined as follows.
The geometry of the spin distribution near a saddle point allows us to compute the mean 
orientation of the spin around the saddle point.
Let us define $\hat \theta$ the flip angle so that 
\begin{equation}
\cos \hat \theta(\mathbf{r}) =\frac{\mathbf{L}(\mathbf{r}).\mathbf{e}_{z}}{||\mathbf{L}(\mathbf{r})||}\,.
\label{eq:deforientation-profile}
\end{equation}

In turn, the shape of the density profile in the vicinity of  the most likely skeleton-like saddle point
(as defined by equation~\ref{eq:defskelsaddle}),
together with an extension of the Press-Schecher theory involving the peak background split 
allows us to estimate the typical mass of dark halos around the same saddle point.

Indeed, the local mass distribution of halos is expected to vary along the large scale 
filament due to  changes in the underlying  long-wave density. In the linear regime,
the typical density near the end points of the filament, where it joins
the protoclusters,  exceeds the typical density near the saddle point
by a factor of two (\cite{Pogosyanetal1998}). At epochs 
before the whole filamentary structure has collapsed, this
leads to a shift in hierarchy of the forming halos towards larger masses
near  filaments end points (the clusters) relative to the filament middle point (the saddle). 
This can be easily understood using the formalism of barrier crossing
(e.g. \cite{Bondetal1991}),
which associates the density of objects of a given mass to
the statistics for the
random walk of halo density,  as the field is  
smoothed with decreasing filter sizes. 
%
%

\begin{figure}
\begin{center}
\includegraphics[width=0.45\columnwidth]{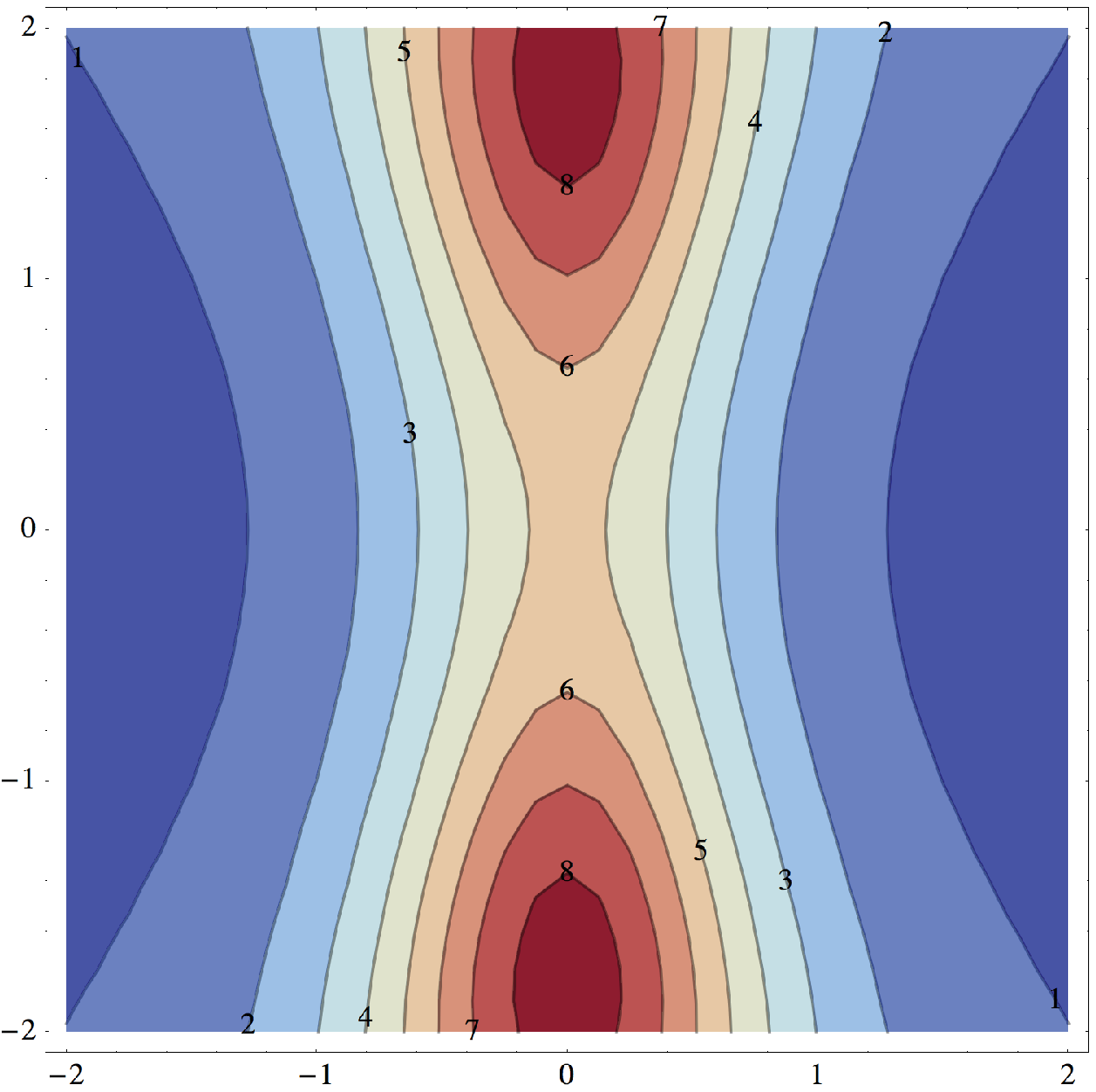}
\includegraphics[width=0.45\columnwidth]{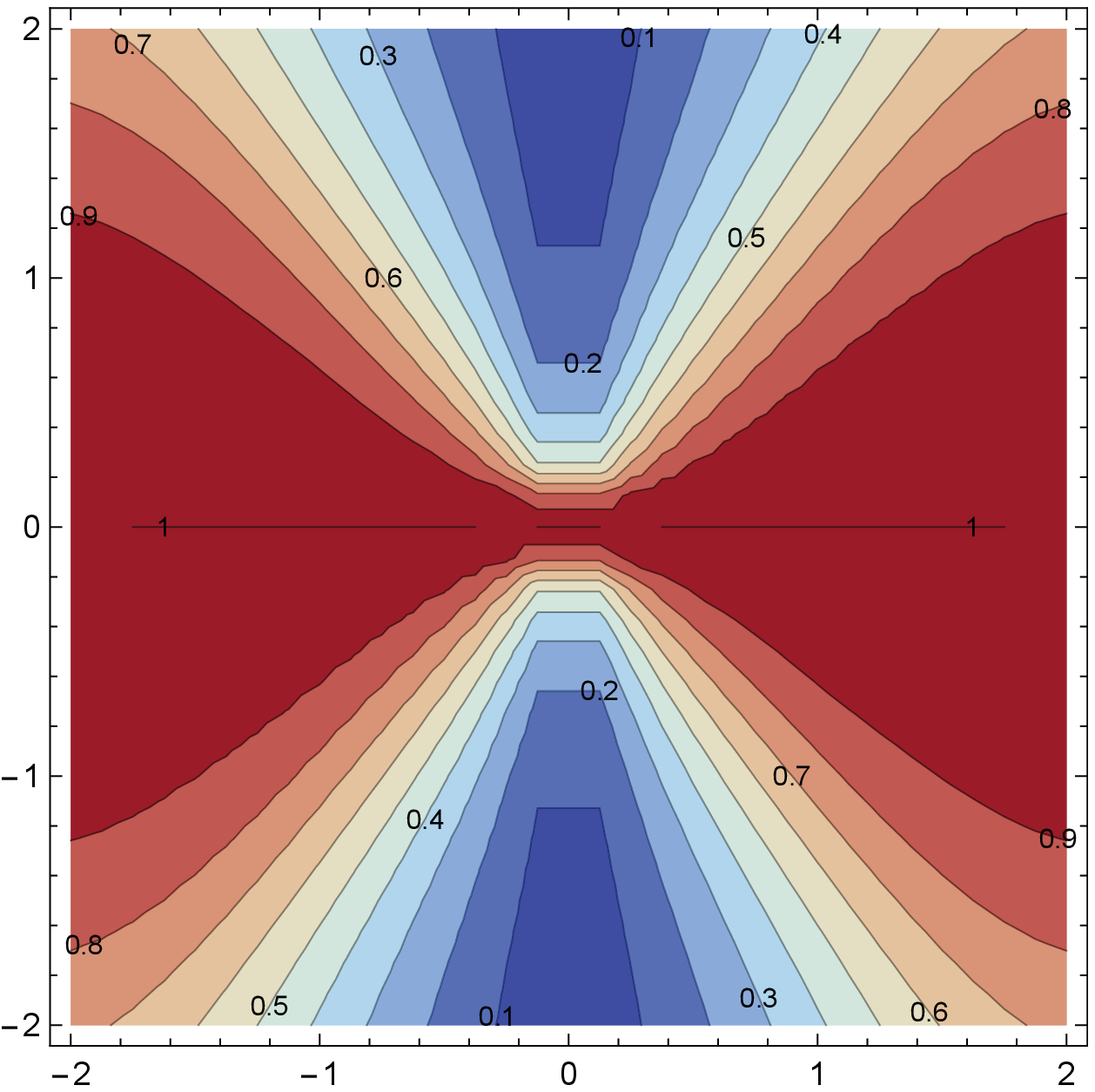}
\caption{{\sl Left:} logarithmic cross section of  $M_p(r,z)$  along the most likely (vertical) filament (in units of $10^{12} M_\odot$).
{\sl Right:} corresponding cross section of $\langle\cos \hat \theta\,\rangle(r,z)$.
The mass of halos increases towards the nodes, while the spin flips.
 }
\label{fig-mass-angle-profile}
\end{center}
\end{figure}

\begin{figure}
\center{
\includegraphics[width=0.7\textwidth]{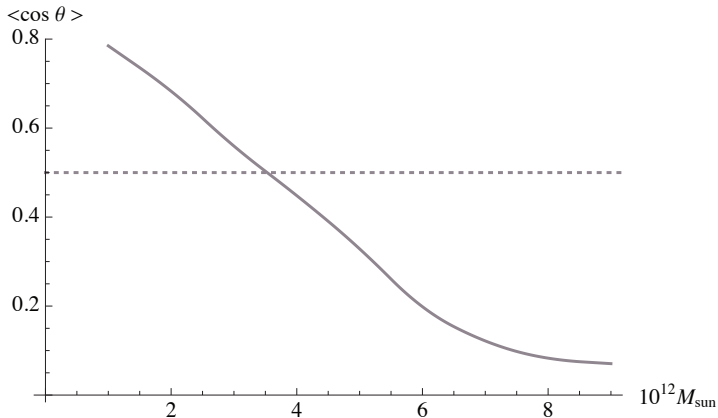}}
\caption{Mean alignment between spin and filament as a function of mass for a filament smoothing scale of 5 Mpc$/h$.  
The  spin flip transition mass is around $4\, 10^{12} M_\odot$.
}
\label{fig:align-mass}
\end{figure}

%
Given the  
Peacock-Heavens (\cite{PH1990})
approximation,
the number density of dark halos in the interval $[M,M+{\rm d}M]$ is
\begin{equation}
\frac{d n(M)}{d M} d M = \frac{\rho}{M}
f(\sigma^2,\delta_c) d \ln \sigma^2\,,
\end{equation}
where $f(\sigma^2,\delta_c) $ is given by the function
\begin{align} \label{eq:deff}
f(\sigma^2,\delta_c) = \exp\left( \frac{1}{\Gamma} \int_0^{\sigma^2}
\frac{d s^\prime}{s^\prime} \ln p(s^\prime,\delta_c) \right)   
 \left( - \sigma^2 \frac{d p(\sigma^2,\delta_c)}{d \sigma^2}
- \frac{1}{\Gamma} p(\sigma^2,\delta_c) \ln p(\sigma^2,\delta_c)\right)\nonumber\,.
\end{align}
Here $\sigma^2$ is the variance of the density fluctuations smoothed 
at the scale corresponding to $M$ and 
$p(\sigma^2,\delta_c)\equiv {1}/{2} \left(1+
\mathrm{erf}(\delta_c/\sqrt{2}\sigma) \right) $ is the probability of
a Gaussian process with variance $\sigma^2$ to yield value below
some critical threshold $\delta_c$. Here $\Gamma \approx 4$. 
The overall mass distribution of halos is well described by the choice
$\delta_{c,0}={3}/{5}\left({3\pi}/{2}\right)^{2/3}=1.686$,
motivated by the spherical collapse model.
When halos form on top of  a large scale structure background, however,
the long-wave over-density $\overline \delta(z)$ adds to  the over-density in the proto-halo
peaks.  The effect on halo mass distribution, in this peak-background split
approach, can be approximated as a  {\sl shifted} threshold 
$\delta_c(z,Z) = 1.686 - \overline \delta(z,Z)$ for halo formation as a function of the curvilinear coordinate
$z$ along the filament and redshift $Z$.
The corresponding shift in mass can be characterized by the dependence on the threshold of
$M_*(\delta_c)$, defined as $\sigma_*(M_*) = \delta_c$, or of the mass
$M_p(\delta_c)$ that corresponds to the peak of $f(\sigma^2,\delta_c)$,
i.e. the variance 
$
\sigma^2_p(z)\equiv \underset{\sigma^2} {\mathrm{argmax}}\Big( f(\sigma^2,\delta_c(z)) \Big)  \,.\label{eq:defsigmap}
$
When large scale structures are considered as
fixed background,  
the variance of the relevant 
small scale density fluctuations
that are responsible for object formation is reduced, approximately as
$
\sigma^2 \approx \sigma^2(M) - \sigma^2(M_{\rm LSS})
\label{eq:sigmaMcorrected}
$
where $\sigma^2(M_{\rm LSS})$  is the unconstrained large-scale density variance.
This correction 
becomes important, truncating
the mass hierarchy at $M_{\rm LSS}$, whenever large scale structures are themselves
non-linear.
Here we choose $\sigma_8=0.8$,  redshift zero,
use the value for mass in a $8 h^{-1}$ Mpc
comoving sphere for the best-fit cosmological mass density,
and approximate the spectrum with a power law of index $n=-2$,
which allows to solve for 
$M(\sigma)$  as
\begin{equation}
M(\sigma,Z) = 2.6 \times 10^{14} M_\odot 
\left(\frac{\sigma^2 + \sigma^2(M_{\rm LSS})}{\sigma_8^2 D(Z)^2} \right)^{-\frac{3}{n+3}}\,. \label{eq:defMsigmaZ}
\end{equation}
We consider filaments defined with $R=5 h^{-1}$Mpc Gaussian smoothing.
Then, in addition to a spin orientation map around the saddle point, one can establish a mass map directly from the density map by means of the $M_p(\delta)$ relation. A cut of those two maps is displayed in Fig.~\ref{fig-mass-angle-profile}. 
The spin flips towards the nodes, while  mass increases. In each point of the vicinity of the saddle point, the mass and spin orientation are known so that one can do an histogram and plot the mean orientation as a function of the mass, see Fig.~\ref{fig:align-mass}.
The 3D transition mass for spin flip (i.e. $\cos \hat \theta=0.5$) is clearly of the order ${\cal M}^{\rm 3D}_{\rm tr} \approx 5\, 10^{12} M_\odot$ 
This mass is in good agreement with the transition  mass found in \cite{codisetal12}.
%
\section{Discussion}
\label{sec:conclusion}

Tidal torque theory was revisited while focussing on an anisotropic peak background split
in the vicinity of a saddle point.  Such point process captures the point-symmetric multipolar
geometry of a typical filament embedded in a given wall (\cite{Pogosyanetal1998}). 
The induced  misalignment between the tidal tensor and the Hessian   
simply  explains  the surrounding  transverse and 
longitudinal point reflection-symmetric geometry of spin distribution near filaments.
It predicts in particular that less massive galaxies have their spin parallel to the filament,
while more massive ones have their spin in the azimuthal direction.
The corresponding  transition masses  (${\cal M}^{\rm 2/3D}_{\rm tr}$, corresponding resp. to maximal alignment and flip, 
see Fig~\ref{fig:spin4pi})  follows from this 
geometry, together with their scaling with the mass of non linearity, as  observed in simulation.
The neighborhood of a  {\sl given  unique}
typical saddle point was considered  as a proxy for the behaviour within a 
Gaussian random field. It is shown elsewhere (Pichon et al. {\it in prep.}) that it holds statistically.

One of the interesting feature of this {\sl Lagrangian} framework is that it captures 
naturally the arguably 
non linear {\sl Eulerian} process of spin flip via mergers.
Recently, \cite{laigle2014} showed 
 that angular momentum generation of halos is captured  via the secondary advection of vorticity which was generated by the formation of  filaments.  These two (Eulerian versus Lagrangian)  descriptions are the two sides of the same coin. 
 The mapping between the two descriptions requires a reversible time integrator, such as the Zeldovitch approximation. 
In effect,  the geometry of the saddle provides a natural `metric' 
(the local frame as defined by the Hessian at that saddle point) relative to
 which  the dynamical evolution of dark halos along filaments can be predicted.
 For instance, from Eq.~(\ref{eq:defL3Dsol}) we can  compute the loci,  along the filament, of maximum angular momentum advection.
They characterize the most active regions in the cosmic web for galactic spin up. The  argument sketched in Section 3  allows us to  assign the corresponding redshift dependent spin-up {\sl mass}, and its  evolution with redshift.
It should have an observational signature in terms of the cosmic evolution of the  SFR, as it
  corresponds to efficient pristine  cold and dense gas accretion, 
  which in turn induces  steady star formation.

\vskip 0.5cm
{\sl  This work is partially supported by grant ANR-13-BS05-0005 of the french ANR. CP thanks D. Lynden-Bell for encouragement.}

\end{document}